\documentclass[aps,prb,twocolumn,groupedaddress]{revtex4}
\usepackage{amsmath,amssymb,bm,graphicx,dsfont,color}

\makeatletter

\newcommand{\Rmnum}[1]{\expandafter\@slowromancap\romannumeral #1@}
\makeatother

\newcommand{\mbz}{{\mathbb{Z}}}

\begin{document}
\title{Beyond Band Insulators: Topology of Semi-metals and Interacting Phases}
\author{Ari M. Turner}
\affiliation{Institute for Theoretical Physics, University of Amsterdam,
Science Park 904, P.O. Box 94485, 1090 GL Amsterdam, The Netherlands}
\author{Ashvin Vishwanath}
\affiliation{Department of Physics, University of California, Berkeley, CA 94720, USA}
\date{\today}
\maketitle
The recent developments in topological insulators have highlighted the important role of topology in defining phases of matter (for a concise summary, see Chapter 1 in this volume). Topological insulators are classified according to the qualitative properties of their bulk band structure; the physical manifestations of this band topology  are special surface states\cite{KaneHasan} and quantized response functions\cite{QiZhang}.  A complete classification of free fermion topological insulators that are well defined in the presence of disorder has been achieved\cite{Ludwig,Kitaev}, and their basic properties are well understood. However, if we loosen these restriction to look beyond insulating states of free fermions,  do new phases appear? For example, are there topological phases of free fermions in the absence of an energy gap? Are there new gapped topological phases of interacting particles, for which a band structure based description does not exist? In this chapter we will discuss recent progress on both generalizations of topological insulators.

In {\bf Part 1} we focus on topological semimetals, which are essentially free fermion phases where band topology can be defined even though the energy gap closes at certain points in the Brillouin zone. We will focus mainly on the three dimensional Weyl semimetal and its unusual surface states that take the form of `Fermi arcs' \cite{Wan} as well as its topological response, which is closely related to the Adler-Bell-Jackiw (ABJ) \cite{ABJ} anomaly. The low energy excitations of this state are modeled by the Weyl equation of particle physics.  The Weyl equation was originally believed to describe neutrinos, but later discarded as the low-energy description of neutrinos
following the discovery of neutrino mass.
 If realized in solids, this would provide the first physical instance of a Weyl fermion. We discuss candidate materials, including the pyrochlore iridates\cite{Wan} and heterostructures of topological insulators and ferromagnets\cite{Burkov}. We also discuss the generalization of the idea of topological semimetals to include other instances including nodal superconductors.

In {\bf Part 2}, we will discuss the effect of interactions. First, we discuss the possibility that two phases deemed different from the free fermion perspective can merge in the presence of interactions. This reduces the possible set of topological phases. Next, we discuss physical properties of gapped topological phases in 1D and their classification\cite{PollmannBoson,ChenOneD,Schuch,Fermions}. These include new topological phases of bosons (or spins), which are only possible in the presence of interactions. Finally we discuss topological phases in two and higher dimensions. A surprising recent development is the identification of topological phases of bosons that have no topological order\cite{KitaevUnpublished,Chencoho2011}. These truly generalize the concept of the free fermion topological insulators to the interacting regime. We describe properties of some of these phases in 2D, including a phase where the thermal Hall conductivity is quantized to multiples of 8 times the thermal quantum\cite{KitaevUnpublished}, and integer quantum Hall phases of bosons where only the even integer Hall plateaus are realized\cite{LuAv,SenthilLevin,WenSU(2)}. Possible physical realizations of these states are also discussed.  We end with a brief discussion of topological phases with topological order, including fractional topological insulators.

\section{Part 1: Topological Semimetals}
A perturbation does not significantly change the properties of a fully gapped insulator, hence readily allowing us to define robust topological properties. Is it possible for a gapless system to have a defining topological feature? How can one distinguish phases of a system with gapless degrees of freedom that can
be rearranged in many ways?

We will consider a particular example of a gapless system, a Weyl semimetal,
as an example. There are three answers to the question of how this state
can be topologically characterized despite the absence of a complete gap: in momentum space, the gapless points
are ``topologically protected" by the behavior of the band structure
on a surface enclosing each point\cite{Volovik,Horava}.  There are topologically protected surface states, as in 3D topological insulators,  which can never be realized in a purely two dimensional system.\cite{Wan}. If a field is applied, there is a response whose value depends only on the location of the gapless points,
but on no other details of the band structure (and so is topological).

In Weyl semimetals, the valence and conduction bands touch at points in the Brillouin zone. A key requirement is that the bands are individually non-degenerate. This requires that either time reversal symmetry or inversion symmetry (parity) are broken. For simplicity, we assume time reversal symmetry is broken, but inversion is preserved.  In practice this is achieved in solids with some kind of magnetic order. 
Consider the following Hamiltonian\cite{Ran,David}, a simple two-band model where the $2\times 2$ Pauli matrices refer to the two bands:
\begin{eqnarray}
H&=&-\sum_k \left [2t_x(\cos k_x-\cos k_0)+m(2-\cos k_x-\cos k_y)\right ] \sigma_x \nonumber\\&&+2t_y\sin k_y\sigma_y +2t_z\sin k_z\sigma_z
\end{eqnarray}
We assume the $\sigma$ transforms like angular momentum under inversion and time reversal, which implies the above Hamiltonian is invariant under the former, but not the latter symmetry. It is readily seen to possess nodes.
At $\mathbf{k}=\pm k_0 \bm{\hat{x}}$, and only there (if $m$ is sufficiently
large), the gap closes.
This will turn out not to be due to symmetry, but to topology. Even if all symmetries
are broken (time reversal already is) a gap will not open. The
gapless points can move around but we will see that they can only disappear if
they meet one another, at which point they can annihilate. This is a key difference from the analogous dispersion in 2D (e.g. graphene) which may be gapped by breaking appropriate symmetries.

If the bands are filled with fermions right up to the nodes, then this is a semimetal with vanishing density of states at the Fermi energy.
We expand the Hamiltonian around the gapless points setting ${\bf p_\pm}=(\pm k_x\mp k_0,k_y,k_z)$, and assuming $|{\bf p}_{\pm}|\ll k_0$:
\begin{eqnarray}
H_\pm&=& v_x[p_\pm]_x \sigma_x +v_y[p_{\pm}]_y\sigma_y+v_z[p_\pm]_z\sigma_z\\
E^2 &=& [{\bf v\cdot p}_\pm]^2,
\end{eqnarray}
where $v_x=2t_x\sin k_0$, $v_{y,z}=-2t_{y,z}$. This is closely related to  Weyl's equation from particle physics:
$$
H^\pm_{\rm Weyl}=\pm c{\bf p}\cdot \sigma
$$
which reduced the $4\times 4$ matrices of the Dirac equation to $2\times 2$ Pauli matrices. Since there is no `fourth' Pauli matrix that anti commutes with the other three, one cannot add a mass term to the Weyl equation. Thus, it describes a massless fermion with a single helicity, either right
or left-handed, which corresponds to the sign $\kappa=\pm1$ before the equation. The velocity $\mathbf{v}=\pm c\bm{\sigma}$, is either along or opposite to the spin depending on the helicity. In a 3D lattice model, Weyl points always come in pairs of opposite helicity;  this is the fermion doubling theorem\cite{NielsenNinomiya}, which is justified below.

The low-energy theory of a Weyl point is in general an anisotropic
version of the Weyl equation,
\begin{equation}
H= \sum_i v_i (\bm{\hat{n}}_i\cdot \mathbf{p})\sigma_i +v_0 (\bm{\hat{n}_0}\cdot \mathbf{p}) \mathds{1}.
\end{equation}
The three velocities $v_i$ describe the anisotropy of the Weyl point,
and the $\bm{\hat{n}}_i$'s are the principal directions, which are linearly independent but not necessarily orthogonal. (We have chosen
the basis for the Pauli matrices such that they match the principal
directions.) The helicity is given by $\kappa={\rm sign}[\bm{\hat{n}}_1\cdot(\bm{\hat{n}}_2\times\bm{\hat{n}}_3)]$ (the sign in front of the Weyl equation after
transforming to coordinates where the velocity is isotropic).

\subsection{Topological Properties of Weyl Semi-metals}
\subsubsection{Stability}
A well known argument about level crossing in quantum mechanics \cite{vonNeumann} can be applied to argue that Weyl nodes are stable to small perturbations regardless of symmetry \cite{Herring}. Consider a pair of energy levels (say two bands) that approach each other. We would like to understand when the crossing of these energy levels (band touching) is allowed. We can write a general Hamiltonian for these two levels as: $H=\left(\begin{array}{cc} \delta E & \psi_1+i\psi_2 \\ \psi_1-i\psi_2 &-\delta E\end{array}\right) $, ignoring the overall zero of energy. The energy splitting then is $\Delta E=\pm \sqrt{\delta E^2 +\psi_1^2+\psi_2^2}$. For $\Delta E=0$ we need to individually tune each of the three real numbers to zero. This gives three equations, which in general needs three variables for a solution. The three dimensional Brillouin zone provides three parameters which in principle can be tuned to find a node. While this does not guarantee a node, once such a solution is found, a perturbation which changes the Hamiltonian slightly only shifts the location of the solution i.e. moves the Weyl node in crystal momentum.   Note, this does not tell us the energy of the Weyl node -- in some cases it is fixed at the chemical potential from other considerations. This is a convenient situation to discuss, but several of the results we mention below do not strictly require this condition. Later we will discuss how Dirac points
can be stabilized in lower dimensions as well, as in graphene, provided that some
symmetries are present.

\subsubsection{Magnetic Monopoles in Momentum Space}
We will now show that the Weyl points' stability is connected to an integer-valued topological index.   That
is the first sense in which a Weyl semimetal is a topological state. There is a conservation law of the net charge of the Weyl points.  That is, a single Weyl point cannot disappear on its own. However, on changing parameters in the Hamiltonian
the Weyl points can move around and eventually pairs with opposite signs can annihilate, meaning
that they first merge and then the bands separate.  In order for Weyl
points to have well-defined coordinates, one needs crystal momentum
to be a well-defined quantum number. Therefore, unlike topological insulators which are well defined even in the presence of disorder, here we will assume crystalline translation symmetry to sharply define this phase.

The Weyl nodes are readily shown to be associated with a quantized Berry flux of $2\pi \kappa$  (where $\kappa=\pm 1$ is the chirality). The Berry flux $\bf B$ is defined as:
\begin{eqnarray}
{\bf \mathcal A}(\mathbf{k})&=& -i\sum_{n,\mathrm{occupied}}\langle u_{n,\mathbf{k}}|{\bf \nabla}_{\mathbf{k}}| u_{n,\mathbf{k}}\rangle,\\
\mathbf{B}(\mathbf{k})&=&{\bf \nabla}_{\mathbf{k}}\times {\bf \mathcal A}.
\label{eq:vecpotential}
\end{eqnarray}
The Berry flux of a Weyl point is readily found from the following simple expression
for the wave function of filled bands near the Weyl point:
\begin{equation}
\psi(\theta,\phi)=\left(\begin{array}{c} \sin\frac{\theta}{2}\\-\cos\frac{\theta}{2}e^{i\phi}\end{array}\right),
\end{equation}
where the momentum coordinates relative to the Weyl point are expressed in terms of the angle: $\theta$ and $\phi$. The Berry phase of this spinor is the same
as the Berry phase of a spin 1/2 object in a field;  one can then determine the
flux through a small sphere about the Weyl point; it is $\pm 2\pi$.  The Berry flux through any surface containing it is the same, thanks to Gauss's law.  Thus a Weyl point may be regarded as a magnetic monopole, since its flux corresponds to
the quantized magnetic charge of monopoles.

Now the stability of Weyl points can be connected to Gauss's law:  a Gaussian surface surrounding the Weyl point detects
its charge, preventing it from
disappearing surreptitiously.  It can only
disappear after an oppositely charged monopole goes through the surface, and later
annihilates with it.
Also, the net charge of all the Weyl points in the Brillouin zone has to be zero (which is seen by
taking a Gauss-law surface that goes around the whole Brillouin zone).
Thus, the minimum number of Weyl points (if there are some) is two, and they
have to have the opposite chirality, as in the model above.  This is the proof of the fermion doubling theorem.

Ideally, to isolate effects of the Weyl points, it is best that they lie right at
the Fermi energy, and that there are no Fermi surfaces (whose contributions to
response properties would drown out signals from the Weyl points). Now it might
seem that it would be necessary to have a very special material and doping
to ensure that the Weyl points are at the same energy and that this coincides with the Fermi energy.  However, a symmetry can ensure the former, and stoichiometric filling, via Luttinger's theorem, the latter. Later examples of materials that have been proposed are given.

Let us compare this approach to identifying band topology with the situation in classifying topological insulators.  The filled bands of a regular topological insulator gives a continuous map from the Brillouin zone to the relevant target space, and one classifies topologically distinct maps.  In a semimetal, the map isn't continuous since the occupied states in the vicinity of a gapless point change discontinuously.  However, a topological property can still be identified
by considering a {\em subspace} of the Brillouin zone that avoids the gapless point,
in this case a two-dimensional closed surface surrounding the Weyl point.

\subsubsection{Topological Surface States}
Now we will look at the surface states of this model.
How can surface states be protected in a gapless system? What keeps the states
from hybridizing with the bulk states?  The answer is translational symmetry:
if the surface momentum is conserved, surface states at the Fermi energy
can be stable at any
momenta where there are no bulk states (at the same energy).
For example, if the Fermi energy is tuned to the Weyl point in the bulk, the bulk states at this energy are only at the projections
of the Weyl points to the surface Brillouin zone. Everywhere else, surface states are well defined because they cannot decay into bulk states at the same energy and momentum. As the Weyl point momenta are approached, the surface states penetrate deeper into the bulk.

\begin{figure}
\includegraphics[width=.45\textwidth]{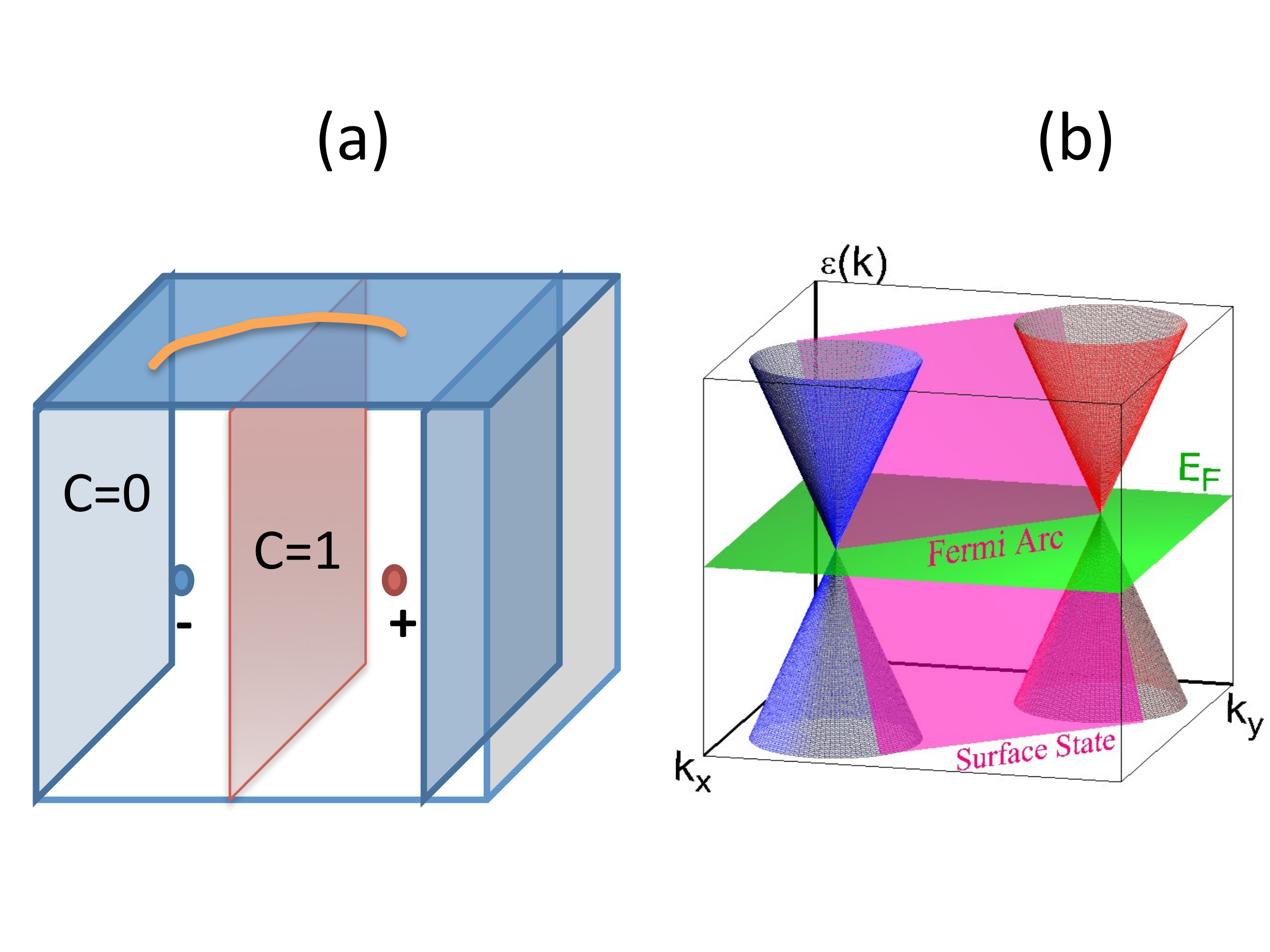}
\caption{\label{fig:sandwich} Surface states of a Weyl semimetal. a)The surface
states of a Weyl semimetal form an arc connecting the projections of the two Weyl points to one another. Points on the Fermi arc can be understood
as the edge state of a two-dimensional
insulator, and the Berry flux of the Weyl points ensures that either
the insulators represented by the red or blue planes are integer Hall
states, which have edge modes. Here we depict the former. b) A graph of the dispersion of the surface
modes (the pink plane) and how this joins to the bulk states (represented by solid red and blue cones)\cite{Wan}.}
\end{figure}

It turns out (see Figure \ref{fig:sandwich}) that surface states form
 an \emph{arc} at the Fermi energy, instead of a closed
curve.
Our usual intuition is that Fermi surfaces form a closed loop since they are defined by the intersection of the dispersion surface $\epsilon=\epsilon(k_x,k_y)$
with the Fermi level $\epsilon=0$. If the Fermi surface is an arc, which `side' corresponds to occupied states?

The key to answering these questions is that surface states are not well defined across the entire Brillouin zone, due to the presence of bulk states as can be seen from
Figure \ref{fig:sandwich}b. The bulk states intervene and allow for the non-intuitive behavior, that is impossible to obtain in a two dimensional system, and in this particular case, also impossible on the surface of a fully gapped insulator.  For example, the Fermi arcs end at the gap nodes where the surface mode is no longer well-defined. Furthermore, the surface representing the dispersion of the surface-states joins up with the bulk states (the blue cones) at positive and negative energies.

Intuitively, we may imagine a complete Fermi surface being present when we begin with  a thin sample of Weyl semimetal. On increasing the thickness, complementary parts of the Fermi surface get attached to the top and bottom surfaces, resulting in arcs. Thus the Fermi surface can be completed by combining the arcs living
on the opposite side of the sample. The surface states are obtained by mapping three dimensional Weyl semimetals to Chern insulators in two dimensions as we now explain.

{\em Semimetals and Topological Insulators -- Relations between dimensions}
Topological semi-metals inherit the topology of an insulator of a lower dimension.
The spherical surface enclosing a Weyl node used above to detect the Berry monopole has another interpretation:
it describes the band structure of an insulator with no symmetries (class A) in two-dimensions.  That is because a 2D insulator is described by a matrix-function $H(k_x,k_y)$
that has a gap everywhere; we only need to trade the Brillouin zone torus
for a sphere in this case. The Hamiltonian matrices become gapped because the surface
avoids the node. These maps are classified by the Chern number, which is identical to the net Berry flux of the magnetic monopole, given by the integral $\int \frac{d^2\mathbf{k}}{(2\pi)^2}{\mathbf B} (\mathbf{k})$ where $\mathbf B$ is defined in Eq. \ref{eq:vecpotential}.

This observation explains why there are surface states in the Weyl semimetal.
We will show that the Fermi arcs connect the projections
of Weyl points onto the surface. The starting point is the fact that
any closed surface in momentum space surrounding one of the Weyl points
has a Berry flux through it.  Such a surface therefore defines a two-dimensional insulator with a Hall conductance,
which must have edge states, and for the right geometry these will correspond
to real edge states of the three dimensional crystal.

Let us consider a crystal that is cut parallel to the $xy$-plane.
 For a clean surface, momentum is conserved, and the wave functions can
be taken to have a definite crystal momenta $\bm{k}_\perp=(k_x,k_y)$
in the direction
parallel to the surface.  Let us in particular fix $k_x$, so that
$\psi(x,y,z)=e^{ik_x x} f_{k_x}(y,z)$. The Hamiltonian then
reduces to a two-dimensional problem, different for each given momentum,
\begin{equation}
H_{k_x}f_{k_x}(y,z)=\epsilon f_{k_x}(y,z)
\end{equation}
This describes a 2D system with an edge, corresponding to the top surface
of the original semimetal.
If two cross-sections contain a monopole between them, then there is a non-zero
net flux through the two of them.  Thus the Chern numbers differ by $1$,
and so one of the two systems at least is a 2D quantum Hall state. For
example, in Fig. \ref{fig:sandwich}a the planes in between the two monopoles
all have Chern number 1.
Thus there is an edge state for each fixed value of $k_x$.  Putting
all these edge states together, we find that there is a Fermi \emph{arc} connecting
the two Weyl points.

That completes the argument for the surface states.
There is another connection between Weyl semimetals and topological insulators in a different dimension: this one appears on {\em increasing} the dimension and considering 4+1D topological insulators (which exist in the absence of any symmetry). The boundary states of these phases correspond to Weyl fermions, i.e.  a \emph{single} Weyl point. Since this is the boundary of a four dimensional bulk, the fermion doubling theorem does not apply. This immediately allows us to draw certain physical conclusions. Disorder on the surface of a 4D topological insulator will not localize the surface states. Hence, if we have a 3D Weyl semimetal with disorder that scatters particles only {\em within} a node, we may conclude that this will {\em not} lead to localization.

\subsubsection{The Chiral Anomaly}
We now consider ``topological responses" of Weyl semimetals.

At first, it may seem that the number of particles at
a given Weyl point has to be conserved, as long as there is translational symmetry.
Yet there is an ``anomaly" that breaks this conservation law.
The conservation of momentum implies that the particles cannot scatter between
the two Weyl points, and hence they can be described by separate equations. The effective description of either
one in an electromagnetic field is
\begin{equation}
H_{R/L}=\mp i\hbar v\psi_{L/R}^\dagger\bm{\sigma}\cdot(\bm{\nabla}-\frac{ie\mathbf{A}}{\hbar})\psi_{L/R},
\end{equation}
where we have assumed the velocity to be isotropic. Response to a low-frequency
electromagnetic field can be calculated from these relativistic equations, using
methods from field theory.
The first step is to add in a cut-off to give a finite answer.  This cut-off does not
respect the conservation law, and so it is found that the charge is not conserved:
\begin{equation}
\frac{\partial}{\partial t}(n_{R/L}(\mathbf{r}))=\pm\frac{e^2}{h^2}\mathbf{E}\cdot\mathbf{B}.
\label{eq:pencils}
\end{equation}
This is known as the Adler-Bell-Jackiw anomaly in field theory.

{\em Condensed matter derivation of the anomaly--}Consider a Weyl semimetal with cross-sectional area $A$ and length $L_z$
in an electric and a magnetic field applied along $z$. The magnetic field causes the electron states to split into a set of Landau levels. As opposed to the two-dimensional case, each Landau level is a one-dimensional mode dispersing along $z$. The zeroth levels are the most interesting. They provide one-dimensional chiral modes with the linear dispersion $\epsilon_{R/L}=\pm vp_z$ for the
right and left-handed species. These modes can be pictured as chiral wires parallel to the magnetic field, with a density per unit area of $B/\Phi_0$ where $\Phi_0=\frac{h}{e}$ is the flux quantum.
 The derivation uses the fact that lowest Landau levels in a 2D Dirac Hamiltonian correspond to an eigenstate of $\sigma_z$, which sets the velocity of $z$ direction propagation. The other modes do not cross through zero,
so they cannot respond much to external fields. Thus, the three dimensional
anomaly reduces essentially to the one-dimensional one.

Now, on  applying an electric field the electrons accelerate
$\dot{p}_z=eE_z$ and so the
Fermi momenta increase as well:
\begin{equation}\dot{p}_{F,L/R}=eE_z,\end{equation}
creating a larger Fermi sea at the right Weyl point, and a smaller one at the left Weyl point.
So effectively the electrons move from the left Weyl point to the right. The rate of
increase gives the exactly correct form for the anomaly
\begin{eqnarray}
\frac{d N_R}{d t}&=&(eE_z)(\frac{L_z}{h})(\frac{AB}{\Phi_0})
\end{eqnarray}
where
$\frac{L_z}{h}$ is the density of states in a chiral mode and $\frac{AB}{\Phi_0}$
is the density of modes.  The number density is obtained from $N_{R/L}$ by dividing
by the volume $AL_z$, and the result agrees with Eq. (\ref{eq:pencils}) for the anomaly.

There is a close connection between the topological surface states and the chiral anomaly.  Consider the following paradox:
In a magnetic field, there are two sets of chiral modes, one traveling along and the other opposite to the field, at different transverse momenta. When a particle in the mode traveling upward reaches the surface, where does it go?  It must get to the opposite node and return back through the bulk, but that is at a different momentum. In fact, it travels along the Fermi arc which connects the projection of the right-handed Weyl point to the left-handed one (the Lorentz force pushes it along the arc).

From the anomaly, it is clear that a single Weyl point is disallowed in a 3D band structure since that would violate particle number conservation. Note, the $\mathbf{E}\cdot\mathbf{B}$ term appears here in an equation of motion, while the same term appears in the effective action for topological insulators \cite{QiZhang}. This connection arises from the fact that the transition between trivial and topological insulators can be understood in terms of Weyl fermions\cite{Murakami,Hosur}.

{\em Anomalous Hall Effect:}  Although this anomaly is a quantized response, it
is hard to measure separately the charge at individual Weyl points. Instead, consider the anomalous Hall effect associated with this simple Weyl semimetal with two Weyl points. The region of the Brillouin zone between the Weyl points can be considered as planes of 2D band structures each with a quantized Hall effect. Thus, the net Hall effect is proportional to the separation between the nodes. Consider creating a pair of Weyl nodes at the center of the Brillouin zone. As the nodes separate, the Hall conductance increases, until they meet at the face of the Brillouin zone and annihilate. At this point the Hall conductance is quantized, in the sense that the system can be considered a layered integer quantum Hall state (Chern insulator) with a layering in the direction ${\bf G}$, the reciprocal lattice vector where the nodes annihilated. Thus the Weyl semimetal is an intermediate state between a trivial insulator and a layered Chern insulator. Likewise, the Fermi arc surface states are understood as the evolution of the surface to ultimately produce the chiral edge state of the layered Chern insulator.
In a general setting\cite{Ran}, the Hall conductivity is given by the vector:\begin{equation}
\mathbf{G}_H=\frac{e^2}{2\pi h}\sum_i \kappa_i \mathbf{k}_i
\label{eq:hall}
\end{equation}
where $\mathbf{k}_i$ is the location of the Weyl points and
$\kappa_i$ is their charge.   In an electric field,
current flows at the rate $\mathbf{J}=\mathbf{E}\times \mathbf{G}_H$. This phenomenon
is a type of quantized response.
The Hall conductivity just depends  on the topological
properties of the Weyl points as well as their locations. \footnote{Sometimes, lattice symmetries can cause this sum to vanish, in which case a different quantity is needed.}
If the latter are determined from a separate experiment, such as ARPES, then a truly quantized response can be extracted. Since this is a gapless state, the accuracy of quantization needs to be theoretically investigated, and is a topic for future work. Certainly, the temperature dependence of deviations from the quantized value vanish, at best, as a power law of temperature given the absence of a finite gap.

Remarkably, this is directly related to the chiral anomaly. Apply both an electric $\mathbf{E}$ and a magnetic $\mathbf{B}$ field to a topological insulator.
Charge will be transferred from the left-handed to the right-handed
Weyl points.  The points are at different momenta, and hence momentum is not conserved. One can compute the rate of change of momentum per unit volume using Eq. (\ref{eq:pencils}):
\begin{equation}
\frac{d\mathbf{p}}{dt}=\frac{e^2}{h^2}\sum_i \kappa_i\hbar\mathbf{k}_i \mathbf{E\cdot B}.
\label{eq:force}
\end{equation}
This is exactly the Hall coefficient, i.e.$\frac{d\mathbf{p}}{dt}=\mathbf{G}_H \mathbf{E}\cdot \mathbf{B}$. This is understood from the fact that electric and magnetic fields acting on a system with an anomalous Hall conductance induce a charge $\rho_H=\mathbf{B}\cdot \mathbf{G}_H$ and current density $\mathbf{J}_H=\mathbf{E}\times \mathbf{G}_H$, which in turn respond with a Lorentz force $\frac{d\mathbf{p}}{dt}=\rho_H \mathbf{E}+\mathbf{J}_H\times \mathbf{B}$, which is equal to $\mathbf{G}_H\mathbf{E}\cdot\mathbf{B}$.  This argument gives an alternative derivation of the Hall conductivity from the anomaly.

\subsection{Topological Semimetals: Generalizations}
As for topological insulators,  topological semimetals can exist in various numbers of dimensions and with different symmetries, and they
all have surface states. In particular, superconducting
systems with nodes can have flat \emph{bands} on their surface\cite{TopSemi,SurfaceStates1,SurfaceStates2,Hook}, which could potentially form interesting phases because interactions would be extremely important.

A topological semimetal is a material whose metallic
behavior has a topological origin; that is, it is not possible
to eliminate gapless points without annihilating them with one another.
Such a material could also have nodes (i.e. band touchings) along loops that are stable, unless the loops
are shrunk down to a point. (We will ignore the energy dispersion along these loops since they can be smoothly deformed to have the same energy).
In general, we define a topological semi-metal in $d$ spatial dimensions as one that has a band touching  along a $q$-dimensional space in the Brillouin zone (where $q\leq d-2$) which is protected by nontrivial band topology of a topological insulator derived from a submanifold of the Brillouin zone surrounding the nodal space.  For a $q$-dimensional node, this topological insulator will be $(d-q-1)$ dimensional, e.g., in $d=3$, a point node will be protected by a sphere surrounding it and a line node will be protected by a circle linked with it.
Here we will restrict attention to those topological states whose symmetries leave the crystal momentum invariant - for example, we will not consider time reversal symmetry, since time reversal changes momentum, and thus it
does not protect nodes at generic points\footnote{Nodes at time reversal invariant momenta may be protected.}.
However, the class A and AIII are both allowed; in the former there are no symmetries, while the latter is based on sublattice symmetry, which leaves the momentum untouched. Invariance of momentum under a defining symmetry allows us to ignore the fact that the
submanifold can be rather different
than a Brillouin zone. (The topology of the manifold is more important when symmetry is present; e.g., on a sphere, no point is invariant under reversing momenta while on a torus there are four invariant points).

{\em Is Graphene a Topological Semimetal?:}  Graphene has two point nodes
in its 2D bulk band structure. However, there is no 1D topological invariant in class A that could be used to protect the degeneracy. Therefore we conclude that real graphene is  a non-topological semimetal. However, if we consider a slightly idealized version of graphene, where the Hamiltonian only involves hopping between opposite sublattices of the honeycomb lattice (such as only nearest neighbor hopping), this turns out to satisfy the criteria for topological semimetal. This version of graphene has a sublattice symmetry, in which changing the sign of amplitudes on one sublattice reverses the sign of the Hamiltonian. This puts it in class AIII, for which there is a 1D topological invariant \cite{Ludwig}. Does this topological protection have other consequences?
Indeed the edge nodes in graphene can be understood from the hidden topological
1D insulators embedded in the band structure. Each line across the
Brillouin zone gives a one-dimensional Hamiltonian.  If a line is in
a non-trivial class, it must have edge states, and in class AIII, these
edge states have strictly zero-energy.
Lines on opposite sides of a node are in different one-dimensional classes, so
if one side has no edge states, the other must have zero-energy states. Indeed, the zero energy state expected e.g.\ at the zig-zag edge\cite{Graphene} of graphene conforms to a topological surface state. However, breaking the sublattice symmetry, for example by changing the potential near the edge, displaces  these zero modes.

The same idea protects \emph{line} nodes in three dimensions in models with sublattice symmetry.  A line
node can be linked with a loop, and if the Hamiltonian on the loop is non-trivial, then
the line node is topological.

{\em Nodal Supercondutors:} As pointed out in Refs.\cite{Volovik, TopSemi,SurfaceStates1,SurfaceStates2,Hook}, this is particularly interesting in the context of superconductors.  In particular, consider a superconductor
with time reversal symmetry $T$.  The superconductor also has a particle-hole
symmetry $\Xi$ (resulting from the redundancy in the Bogolyubov-de Gennes equations),
and the product, $\Xi T$ is a unitary symmetry
that anticommutes with the Hamiltonian, and leaves the crystal momentum $\mathbf{k}$ invariant. Thus, the loop is again classified
by AIII symmetry.  If it is non-trivial, then the line node cannot
be removed. This is the most natural way for the AIII symmetry to arise--the chiral
symmetry could just happen to be an exact symmetry, as for ideal graphene,
but any small farther-neighbor hopping will generically break it.

The definition of a topological semimetal as we formulated it generalizes the
``topological protection" of Weyl
points by their magnetic charge to other systems.  Do the other
properties, surface states and topological responses, come as consequences?

The argument for surface states given in the graphene case and for the Weyl semimetal
certainly generalizes; they are the edge states of $(d-q-1)$-dimensional cross-sections
of the Brillouin zone that have a non-trivial band topology on account of the non-trivial
topology of the nodes.
For example, there are exactly zero-energy
states on the surface of a superconductor with topological node-loops:
For any facet direction that the superconductor is
cut along, the loop of nodes casts a ``shadow" onto the surface
Brillouin zone.  On either the inside or the outside there has to be a state
with energy that is strictly equal to zero.

Indeed, the well known zero bias peaks in d-wave cuprate superconductors are also an example of edge states of a topological semimetal (the d-wave singlet superconductor)\cite{TopSemi}. Applications of this idea \cite{TopSemi,SurfaceStates1} could help determine pairing symmetry  in a nodal superconductor.

Do quantized responses, which we define as ones that depend only on the locations and
topological properties of nodes, generalize as well?  So far there is only
one example, the Hall effect in a Weyl semimetal.  It is ``topological" in the sense that it is determined by
the location of the Weyl nodes and their indices, but does not depend on the wave
functions as e.g. electrical polarizability does.
There is a second example if we allow an ordinary metal to count as topological: Luttinger's theorem. Luttinger's theorem says that
the density of electrons (which plays the role of the response) is proportional
to the volume of the
Fermi surface. (The Fermi surface in an ordinary metal can be regarded as a topologically protected node with $q=d-1$--however, because it is so ordinary we ruled it out
explicitly above by assuming $q\leq d-2$.)

\subsection{Candidate Materials}
While there is no experimental confirmation of a Weyl semimetal phase, several promising proposals exist:

{\em (i) Pyrochlore iridates:}  $A_2Ir_2O_7$, form a sequence of materials (as the element
$A$ is changed) with increasingly strong interactions. $A$ can be taken
to be a Lanthanide element or Yttrium. While some members such as $A=Pr$ are excellent metals, others such as $A=Nd,Sm,Eu\, {\rm and} \,Y$ evolve into poor conductors below a magnetic ordering transition. These may be consistent with an insulating or semi-metallic magnetic ground state\cite{Maeno}.

\begin{figure}
\includegraphics[width=.45\textwidth]{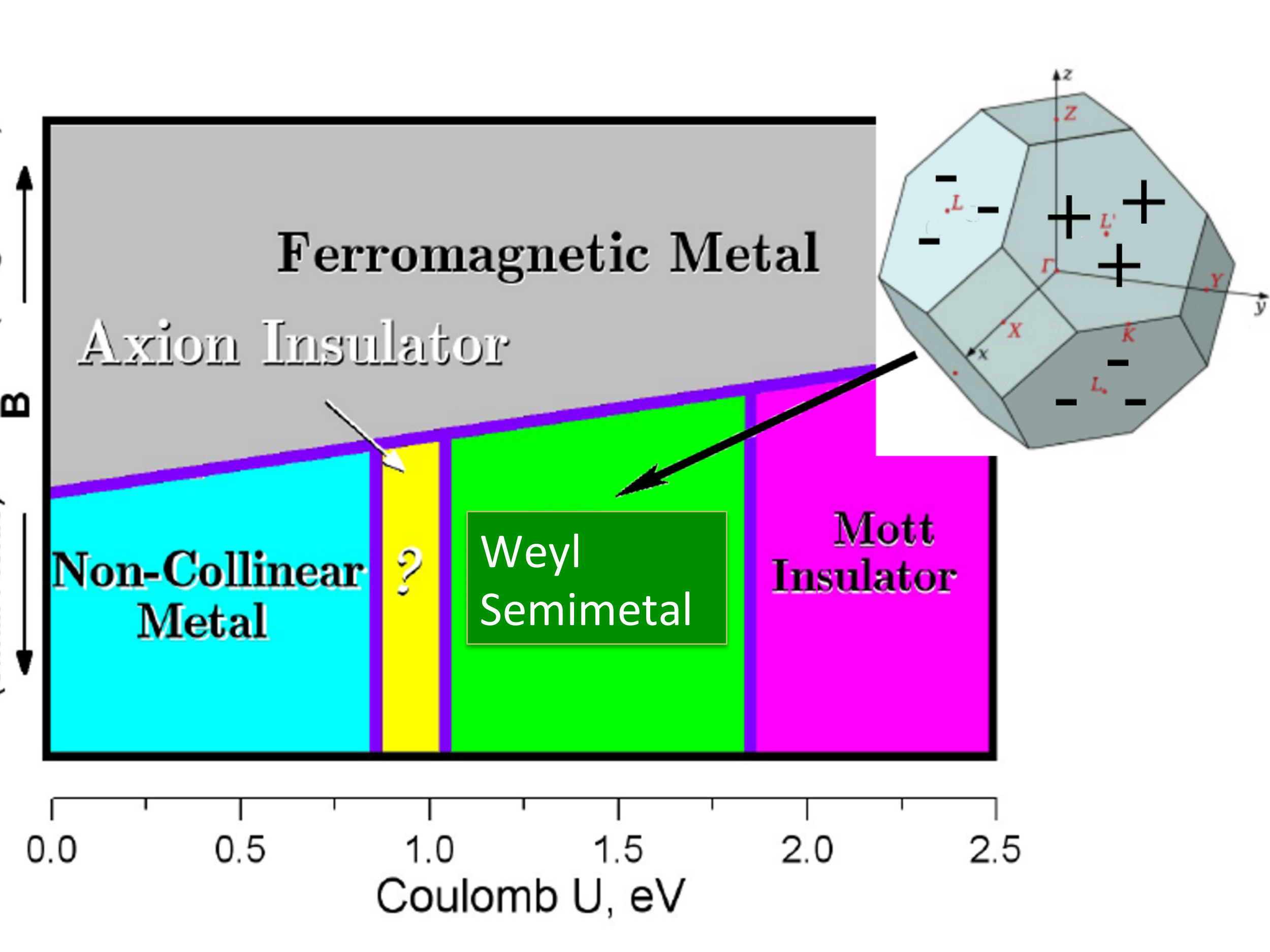}
{\caption{\label{fig:phasediagram} A sketch of the calculated phase diagram of pyrochlore iridates with interaction strength (U) on the horizontal axis. The Weyl semi-metal appears in a regime of intermediate correlations that is believed to be realized in some members of the pyrochlore iridates. It appears as a transition phase between a trivial (Mott) and topological (Axion) insulator. The inset shows the location of the 24 Weyl nodes, labeled by their $\pm$ chirality, as predicted for this cubic system [\onlinecite{Wan}].}}
\end{figure}

Recent ab-initio calculation \cite{Wan} using $LSDA+SO+U$ method predict a magnetic ground state with all-in all-out structure. On varying the on-site Coulomb interactions from strong to weak (see Fig. \ref{fig:phasediagram}), the electronic structure is found to vary.   At very strong interactions, an ordinary magnetically ordered Mott insulator is expected. (This would have no topological properties since strong interactions overwhelm hopping, thus ruling out interesting charge-carrying surface states.)
As the interactions become weaker, the system possibly realizes an ``axion
insulator", which is analogous  to a strong topological insulator
but with inversion rather than time reversal symmetry. A competing state is a magnetic metal. Intermediate between
these is the Weyl semimetal phase. This sequence of phase transitions is natural because an axion insulator requires two
band inversions while a Weyl semimetal requires only one. There were 24 Weyl points (12 of each chirality) predicted that are related to one another by the cubic point-group of this crystal. This implies all 24 Weyl nodes are at the same energy, and at stoichiometry, the chemical potential passes through the nodes.  The anomalous Hall effect cancels out because of the symmetry, but if the material
is strained, there will be a Hall coefficient that is proportional to the amount of the strain\cite{Ran}.
Other effective theories find broadly similar results \cite{Kim,Chen} and related phases have been predicted in spinel osmates \cite{Wan2}. Currently experiments are underway to further investigate these materials.

{\em (ii) TI Based Heterostructures:} Ref.\ \onlinecite{Burkov} showed that layering a topological insulator with a ferromagnet produces a Weyl semimetal with just two Weyl points. Ref.\ \onlinecite{Halasz} showed that it is also possible to make a Weyl semimetal without breaking time-reversal symmetry:
it proposes layering  $HgTe$ and $CdTe$ and then applying an electric field perpendicular to the layers to break \emph{inversion} symmetry. Magnetic doping near the quantum phase transition between a topological and trivial insulator is also expected to open a region of Weyl semimetal\cite{Cho}.

{\em (iii) Spinel $HgCr_2Se_4$} Ab initio calculations of Xu et al.\cite{Xu} on $HgCr_2Se_4$ show that it is likely to realize a pair of Weyl nodes. It is known to be ferromagnetic, thus breaking time reversal symmetry. The Weyl nodes are `doubled' in that they carry Chern number $\pm2$, and the low temperature metallic nature is explained by the presence of other bands crossing the Fermi energy. Nevertheless unusual surface states should be visible.

{\em (iv) Photonic Crystals:} Lu et al.\cite{Lu} showed that light can be given a Weyl dispersion by passing it through a suitable photonic crystal. This scenario includes Weyl line nodes, which imply flat dispersions on the surface of the crystal, i.e., photon states on the surface with zero velocity.

The symmetry properties of crystals that could realize Weyl semi metals are discussed in \cite{Manes} and a  related massless Dirac semimetal was discussed in \cite{Mele}. Given how natural the Weyl semimetal is as the electronic state of a 3D solid, these materials candidates may be just the tip of the iceberg.

\subsection{Other Directions}
In Weyl semimetals, both the surface and the bulk entertain low energy excitations. The role of disorder and interactions on these and their interplay needs much more study. Initial studies have focused on transport. Simple power counting arguments reveal that disorder is irrelevant, unlike in graphene where it is marginally relevant. Hence weak disorder will not induce a finite density of states in a Weyl semimetal. The conductivity displays a more intricate behavior than this simple analysis suggests\cite{Burkov,Hosur1,Hook}. Turning to the effect of interactions, long range Coulomb interactions are marginally irrelevant, and the possibility that they dominate transport was considered in \cite{Hosur1}. In real samples, charged impurities are likely to move the chemical potential away from the Weyl nodes \cite{Hook}. Anomalous magnetotransport  signatures of Weyl nodes were discussed in \cite{Aji,Burkov1,Son1,Son2}, but other smoking gun signatures of the ABJ anomaly would be of great interest. The question of analogous responses in nodal superconductors is also an open problem. Instabilities of the Weyl semi metal state to density wave \cite{Zhang1,Aji1} and superconducting\cite{Sau,Balents2,ChoBandarson} instabilities, as well as the nature of topological defects\cite{Qi1} were also recently studied.

\section{Part 2: Topology of Interacting Phases}

A major theme that saturates the theory of topological insulators
 is ``What are the possible phases of systems that do not
break symmetries?"  More specifically, what are some qualities that distinguish them? We will focus on states characterized by unusual surface modes, but which are otherwise conventional.
Although this ignores interesting topologically ordered states such as fractional Quantum Hall  phases in two dimensions,
it also captures many interesting phases, and is a very natural generalization
of topological insulators, whose surface states (such as Dirac modes
and Majorana fermions) have been measured with photoemission, tunneling, and so on.

\begin{figure}
\includegraphics[width=.45\textwidth]{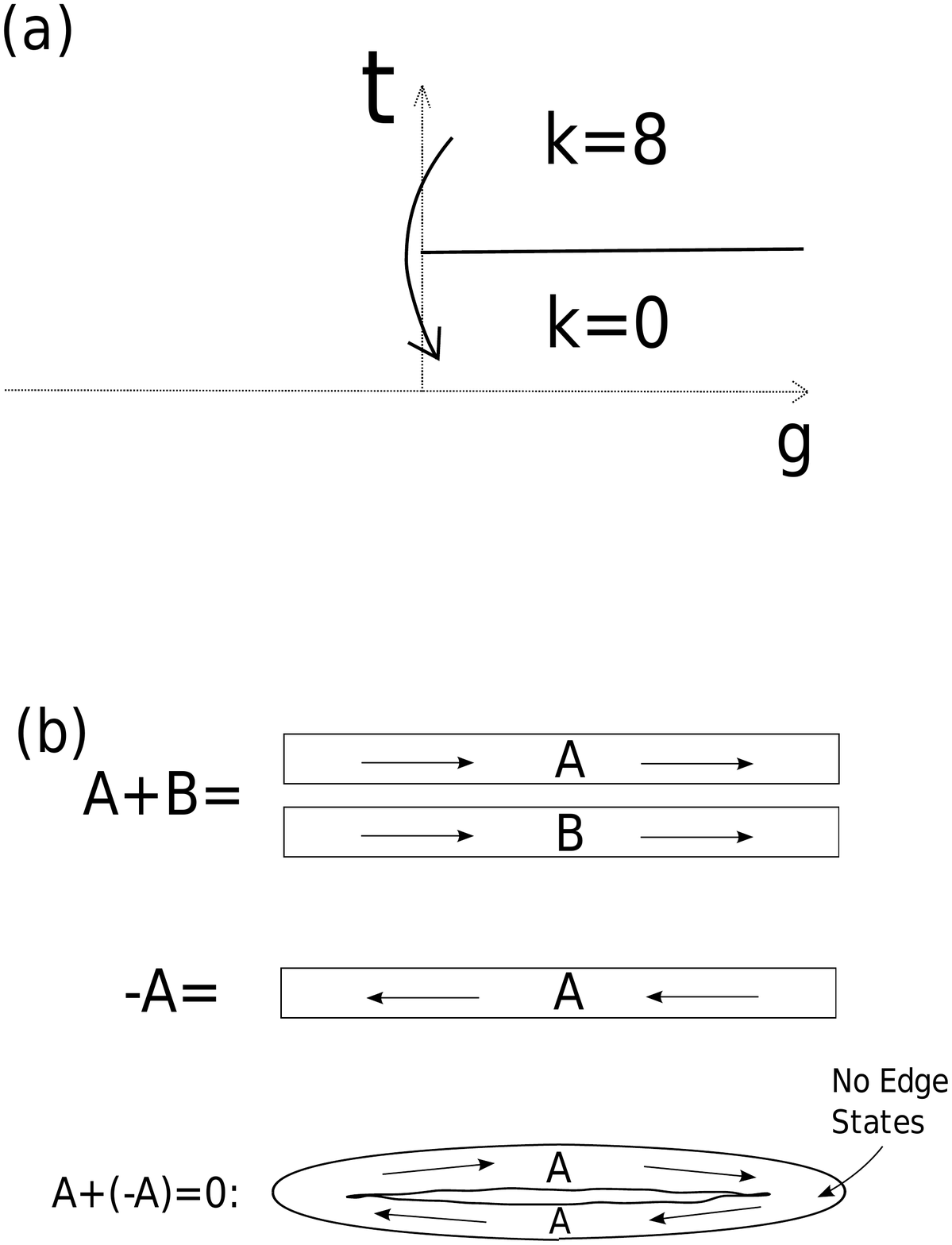}
\caption{\label{fig:sailboat} a) Effects of interactions on classification
of topological insulators.  A schematic phase diagram (based on two of
the phases of Majorana fermion chains) with a
parameter ($t$) related to hopping and an interaction parameter ($g$).  Along
the non-interacting vertical axis, this diagram contains two distinct phases.
When a negative interaction is turned on, the two phases connect to each other.
b) The group operations on topological phases: addition is represented
by layering states, the negative of a state is its mirror image.  The last
figure shows why this is. The arrows are supposed to indicate that
the states may break reflection symmetry.}
\end{figure}

Interactions can
change the classification of topological insulators.
There are two possible effects they can have--they can cause
phases to merge, or they can allow for new phases to exist which could not
exist without interactions.
A surprising example of the former type is that the classification of a certain
type of one-dimensional topological insulator (class BDI) changes\cite{Interaction}: it
is classified
by integers when there are no interactions, so there
are infinitely many phases; when interactions are added,
the classification changes to $\mathbb{Z}_8$, see Fig. \ref{fig:sailboat}.

When one
asks how interactions change the classification of topological insulators, one
means ``how does the connectivity of the regions
in the phase diagram change due to
interactions", not, for example, more involved questions
like how the dynamics of excitations is affected
by scattering or even about the specific form of the phase diagram.
If there is some path joining
one state to another without passing through a discontinuity e.g.\ in the ground
state energy, then these states are defined to be in the same phase.
Imagine adding an axis with a parameter
that
represents the strength of interactions, as in Fig. \ref{fig:sailboat}.
When the interaction is turned on, some distinct phases in the non-interacting
case may merge, similar to how the liquid and vapor phases of water merge at
high pressure. This is exactly what happens in the BDI class of
superconductors: When interactions are turned on, the two phases
that are labelled by $0$ and $8$ join, as in the phase diagram shown, although
in the non-interacting limit  these states
remain separated by a phase transition no matter how many axes are added.

\subsubsection{Phases with Short Range Entanglement}

We will restrict ourselves to a certain type of phases with Short Range Entanglement ({\bf SRE}). Physically, this means that we look at phases with (i) An energy gap in the bulk and (ii) No topological order, i.e. a unique ground state when defined on a closed manifold. At the same time these requirements impose a certain locality condition on the quantum entanglement. More precisely,  we require that the conditional mutual information
of the variables in any three disjoint regions $A,B,C$ tends to zero exponentially as the
distance between the two closest points in $A$ and $B$ tends to infinity.   The conditional mutual information between three sets of variables $A$,$B$,$C$
is defined as
$S(A,B|C):=S(AC)+S(BC)-S(C)-S(ABC)$ where $S$ is the entanglement entropy.
It roughly captures
the correlations between variables $A$ and $B$ after variable
$C$ is measured. In gapped phases with topological order, this yields the topological entanglement entropy\cite{LevinWen}. This differs slightly from the definition by Chen, Gu and Wen\cite{Chencoho2011} of SRE phases, which are further required to be able to be continuously changed into a product state without a phase transition when symmetries are allowed to be broken.

A characteristic of topological states with only SRE
is unusual edge excitations, namely edge states that cannot exist in
a true system of $d-1$ dimensions, but only on the boundary of the
$d$-dimensional topological state.
 For example,
the doubling-theorem says that a chiral mode cannot exist on a discrete
one-dimensional lattice, yet it can be simulated at the edge of a two-dimensional
quantum Hall state.
Another example is
a chain made up of integer spin particles.  As we will discuss, they
can have half-integer spins at the ends,
even though half-integer spins can't be made out
of integer spin particles in a zero-dimensional system.

When long-range-entanglement is allowed, an even more fascinating phenomenon
can occur: anyon excitations in the bulk,
like in the fractional quantum Hall
effect. These states however are much more complicated to
classify, and so we stick to the SRE states.

Part of the interest of SRE phases is that they have an
orderly and interesting structure: they form a group; thus classifying states (such
as the integer quantum Hall states) is tractable, much simpler than classifying phases with anyons in the bulk.
We can give a simple argument for this based on the assumption that SRE phases can be classified entirely by their edge states, whereas Kitaev has given a more general argument\cite{KitaevUnpublished}. (The assumption is correct in one dimension as long as none of the symmetries imposed are spatial symmetries like translation or inversion symmetry, and there are no counterexamples known in two-dimensions.)

The operations
for the group
are shown in Fig. \ref{fig:sailboat}b. The addition operation is obvious:
take two states and put them side-by-side. But it is not so obvious that a state
has an inverse: how is it possible to \emph{cancel out} edge states? Two copies
of a topological insulator cancel one another, because the Dirac points can
be coupled by a scattering term that makes a gap.  A bigger surprise is that
8 copies
of the 1 dimensional Majorana chain can be coupled by quartic interactions
in a way that makes a gap.

The inverse of a phase is its mirror image. To see this, we must show that the state and its mirror image cancel; the argument is illustrated in the last frame of the figure.
In one dimension, for example, take
a closed loop of the state, and flatten it.  The ``ends" are really part of the bulk
of the loop befored it was squashed, so they are gapped.  Therefore
this state has no edge states.
Because topological SRE phases are classified by their edge states, it must
be the trivial state.

Phases with long-range-entanglement on the other hand
do not form a group because they are not classified
by their boundary excitations.  When two layers are combined
as above, the edges can be gapped, but now there
are even more species of anyons in the bulk (one for each layer), and
the state is even farther from being equal to the identity.
Hence long-range-entangled states
do not have as much structure to organize them.

\subsection{One Dimensional Topological Insulators}
Classifying topological
insulators becomes difficult with interactions because there is
no single-particle band structure that allows us to reduce the problem to studying
the topology of
matrix-valued functions on a torus.
 Since it is very complicated to solve a many-body Hamiltonian, we start with
the ground state directly.  In one dimension (with the exception of Majorana
phases of fermions), a symmetry must be imposed in order to produce distinct phases.

Since we cannot take advantage of band structure in the interacting case,
all we have is 1) the many-body
low-energy eigenstates which are localized around the edge,
and 2)the symmetry operations. How can these be combined together
to distinguish different phases?

A model of an interacting topological phase (that is analogous
to a topological insulator)
is the Heisenberg spin chain
of spin 1 atoms, which has the Hamiltonian
\begin{equation}
H=\sum_i \mathbf{S}_i\cdot \mathbf{S}_{i+1}.
\end{equation}
This Hamiltonian has an $SO_3$ symmetry. The bulk of the ground
state has an $SO_3$ symmetry as well (thanks to the Mermin Wagner theorem), while the edge states are spin $\frac{1}{2}$ excitations.

These excitations have been seen in a compound called NINAZ (among others, see also
Refs. \onlinecite{originalAKLT})
that consists of
chains of spin 1's (carried by nickel atoms). Cooling the material
through a structural transition breaks the chains up, creating ends.
Then the signal from electron spin resonance and the temperature-dependence
of the magnetic susceptibility\cite{Granroth} behave as if there are
free spin $\frac{1}{2}$'s making up
the system, even though the atoms all have spin one.

What does it mean to say that the spin one chain has spin $\frac{1}{2}$'s at the end?
First of all, this Hamiltonian has four nearly degenerate ground states, the same as one
would expect from a pair of spin-$\frac{1}{2}$ particles that are
far enough apart that they cannot interact. In order to justify this interpretation,
note that the four ground states are not symmetric under rotations
of the full chain,
$\Sigma(\bm{\hat{n}},\theta)=\prod_j e^{-i\mathbf{S}_j\cdot\bm{\hat{n}}\theta}$. Since
the order parameter in the bulk vanishes, it seems (although
this may be classical intuition at work)
that the asymmetry
must be associated with asymmetric operators at the ends\footnote{An exotic alternative is that there is some asymmetric correlation among all the spins in
the chain simultaneously, but it is not the case.}. This can be tested by applying a magnetic field to just one end.  This lifts the degeneracy partly: afterward the levels
form two doublets, which indicates that there is one spin $\frac{1}{2}$ at the end the field is applied to,
and another at the other end that remains unpolarized.

The bulk is symmetric
under rotations, and only the ends change\footnote{The wave function cannot
be divided into a bulk wave function and an edge wave function, on account of entanglement.  The precise meaning of this statement is that the \emph{reduced density matrix} of the interior of the chain is not affected by applying $\Sigma$.}.
 Therefore, we can write $\Sigma$ as
\begin{equation}
\Sigma(\theta,\bm{\hat{n}})|\alpha\rangle=L(\theta,\bm{\hat{n}})R(\theta,\bm{\hat{n}})|\alpha\rangle, \label{eq:spin}
\end{equation}
for all the ground states $|\alpha\rangle$, where $L(\theta,\hat{n}),R(\theta,\hat{n})$ are some operators on sites
within the correlation length of the ends\footnote{Intuitively, $\Sigma$ is defined by cuting off several spins starting at the end, rotating them, and then correcting the bonds that have been broken.}. These operators
are called the \emph{effective edge symmetries}\footnote{Since there are only four
ground states, this condition does not uniquely define the edge
spin operators.  However, we can enlarge the space of edge states
by supposing that the coupling between spins increases adiabatically over
many spins
to a very large value in the bulk; there will then be many sub-gap states
and Eq. (\ref{eq:edgespin})
can be required to hold for all of them.}.

We can now introduce effective spin operators by defining $L(\theta,\bm{\hat{n}}):=e^{-i\mathbf{s}_L\cdot\bm{\hat{n}}\theta}$ and similarly for $R(\theta,\bm{\hat{n}})$.
We then have
\begin{equation}
\mathbf{S}_{total}=\mathbf{s}_L+\mathbf{s}_R.\label{eq:edgespin}
\end{equation}
This is true \emph{within the space of ground states}.
When we say that the edges have half-integer spins, we mean that they transform
as half-integer spin multiplets under $\mathbf{s}_L$ and $\mathbf{s}_R$.
What is the actual form of these operators?  They are not just the sum of a sequence of $N$ spins at one of the ends (for some
$N$)
because
any eigenvalue of this would have to be an integer. Such a sum is inappropriate
for defining the spin of the edge because it does not commute with the
density matrix of the bulk; measuring this operator breaks bonds from
the $N^\mathrm{th}$ atom, the last whose spin is summed. This alters
the answer
 because it creates nonzero angular momentum
around the $N^\mathrm{th}$ atom.
The correct
operator must be modified to keep the bonds intact.

Now spin chains can be divided into two classes. The eigenstates of $\mathbf{s}_L$
and $\mathbf{s}_R$ come in multiplets.  Either all multiplets are integer or all multiplets are half-integer spins.  Otherwise, one could pair a half-integer state
on the left end with an integer state on the right end (since the two ends are
independent).  Then the net spin is a half-integer which is impossible, since it must match the physical spin $\mathbf{S}_{total}$ and
that is an integer.

The key distinction among spin chains is whether the edge spin is an integer or a half-integer spin; the actual value of the spin is not important.
For example, a state with an integer spin at the end (e.g.,
a spin $2$ Heisenberg chain, which has a spin $1$ at the end) can be adiabatically
transformed into a state without a spin at the end; there will be a level
crossing at some point, but this only has to occur at the ends. When the edge spin changes by a \emph{half}-integer there must be a gapless mode in the bulk at some point.

\emph{General Classification of Spin Chains and Bosonic Systems} --
The idea of fractional edge states can be generalized beyond this special
example. For each symmetry group, there is a definite list of phases associated
with the group. The classification depends only on the abstract type of the symmetry group, and not on what it describes.
For example, the classification applies to systems of mobile bosons, as well as fixed spins. Conservation
of particle number produces a $U(1)$ symmetry $\psi\rightarrow e^{i\theta}\psi$.  If there are two species of particle, then there are two $U(1)$ symmetries.
Alternatively the symmetries can be spatial symmetry with respect to a cross-section:
for example, four
chains with a square cross-section can have the symmetry group $D_4$.

For an ordinary
symmetry-breaking transition, a phase is specified by \emph{the symmetry group}
and the \emph{list of broken symmetries}.

In a topological state no symmetries are broken in the bulk, so the rule is more subtle.
One first observes that the edges might spontaneously break symmetry, even though the bulk does not.
Thus, the topological phases are defined by how the edge states transform
under the symmetries.
Generalizing Eq. (\ref{eq:spin}) to other symmetry groups gives a factorization
\begin{equation}
g_i|\alpha\rangle=L_{g_i}R_{g_i}|\alpha\rangle
\label{eq:otherhand}
\end{equation}
for each element $g_i$ of the group.

We now need a
way to distinguish ``fractional" and ``integer" symmetries $L_g$ for discrete
symmetry groups. We classified the $SO_3$-symmetric phases above (i.e., spin chains) by whether the spins of the ends were integer or half-integer valued--that is, by checking whether the eigenvalues of
$s_{Lz}$ were integers or half-integers.  In the more general case we cannot simply look at the eigenvalues of $L_{g_i}$
because the factorization Eq. (\ref{eq:otherhand})
into $L_{g_i}$ and $R_{g_i}$ is not unique,
thanks to the transformations $L_{g_i}\rightarrow e^{i\alpha} L_{g_i}$, $R_{g_i}\rightarrow e^{-i\alpha} R_{g_i}$.
 This ambiguity makes the eigenvalues of individual $L_{g_i}$'s meaningless,
but it turns out also to be the key to a different approach to defining
fractionalization.

We must consider several symmetries
simultaneously.
The action of the symmetries on the left end of the chain defines
a ``projective representation" of the actual symmetry group.
For each $g_i$, make an arbitrary choice
for the phase of $L_{g_i}$.
The effective symmetries then multiply in a way
that differs from the original group of symmetries by phase factors, called
``factor sets,"
\begin{equation}
L_{g_i}L_{g_j}=e^{i\rho(g_i,g_j)}L_{g_ig_j},
\label{eq:factorset}
\end{equation}
on account of the ambiguity in the phase.
\emph{Topological phases can then be classified by the values of the
factor sets for the edge states}\cite{PollmannBoson,Schuch,ChenOneD}.

Understanding how this works requires some more elaboration.  Although the phase of $L_g$ is ambiguous, there is a certain part of the phase $\rho(g_1,g_2)$ that
cannot be attributed to this ambiguity. This is the part that can be used
to classify the phases.  In fact, the $\rho$'s can be divided into a discrete
number of inequivalent classes, and it is not possible to go from one class
to another without either a discontinuous transition or a point where the
gap closes (so that the edge state operators temporarily lose their meaning and
thus the $\rho$ phases can change).
The following two examples illustrate how discrete phases come about.

\emph{Classifying states with $\mathbb{Z}_2$ and $\mathbb{Z}_2\times \mathbb{Z}_2$
symmetry.}
For a single order $2$ symmetry $g$, it is possible to eliminate the $\rho$-phase
(there is only one)
 by making a different choice for the phase of $L_{g}$ in the first place.
If $L_{g}^2=e^{i\rho}\mathds{1}$,
then redefine $\bar{L}_{g}=e^{-i\rho/2}\mathds{1}$, and the phase has
been eliminated.

For two order two symmetries that commute the phases
cannot necessarily be completely absorbed by redefining
the phases of the $L_g$'s.  Let us call the symmetries $\Sigma_x$
and $\Sigma_z$
and call the effective edge operators
at the left end $L_x$ and $L_z$.
As in the above, $L_x^2$ must be a phase, but this phase can
be absorbed into the definition of $L_x$.
More usefully,
\begin{equation}
\Sigma_x\Sigma_z\Sigma_z^{-1}\Sigma_x^{-1}=\mathds{1}
\end{equation}
which implies
\begin{equation}
L_xL_zL_x^{-1}L_z^{-1}=e^{i\phi}\mathds{1}
\label{eq:PHI}
\end{equation}
because of the phase ambiguity. \emph{This} phase cannot be eliminated
no matter how the phases of the $L$'s are redefined, and
so the value of $\phi$ can be used to discriminate between topological states.
In addition, from the fact that $L_x^2\propto \mathds{1}$ it follows
that $\phi=0$ or $\phi=\pi$, giving two distinct phases of systems
with this $\mathbb{Z}_2\times\mathbb{Z}_2$ symmetry. The $\phi=\pi$ case
is identified as a fractional phase: because the symmetries anticommute, the ends must be two-fold degenerate, like spin-$\frac{1}{2}$s.

One application of this example is to spin chains as discussed above, but
with crystal
anisotropy. Suppose
that the
spin chain has an anisotropy
such that the $x$,$y$ and $z$ components of the spin have
different coupling strengths (i.e. $H=\sum_i J_xS_{i,x}S_{i+1,x}+J_yS_{i,y}S_{i+1,y}+J_zS_{i,z}S_{i+1,z}$).
Then the only remaining
symmetries are $180^\circ$ rotations around the three axes: we will focus on
the rotations around $x$ and $z$, $\Sigma_x$ and $\Sigma_z$ (since the third rotation is just their product).

 We want to be able to distinguish between
integer and half-integer edge-states.  One idea is to characterize a half-integer
spin by the fact that it picks up a minus sign
under a $360^\circ$ rotation.  The phase of the rotation $L_x^2=\pm\mathds{1}$
acting on the edge is not well-defined however.
The characterization
above does distinguish two phases, though.
The two symmetries commute, $\Sigma_x\Sigma_z=\Sigma_z\Sigma_x$ (most pairs of rotations do not commute, but two $180^\circ$
rotations around orthogonal axes do, as one can check by rotating a book around
two axes of symmetry). Thus there are two phases possible.

One can check that $180^\circ$ rotations
around $x$ and $z$ of a spin either commute or anticommute depending
on whether the spin is an integer or half-integer; thus the two possible
phases are the vestiges of the phases of $SO_3$-symmetric spin
chains with integer
and half-integer spins at the ends, respectively; the full $SO_3$ symmetry
is not necessary to distinguish the states.

\emph{General Groups} --
Two states are in the same phase if the factor set $\rho(g_i,g_j)$ is the
same; indeed there is always a path connecting them continuously
in this case, as shown by Ref. \onlinecite{Schuch,ChenOneD}. Thus, the phases for a general
group can be analyzed by solving for all the possible factor sets.
The $\rho$'s must
satisfy a set of simultaneous equations,
\begin{equation}
\rho(g_ig_j,g_k)+\rho(g_i,g_j)=\rho(g_i,g_jg_k)+\rho(g_j,g_k) \mathrm{mod\ }2\pi,
\label{eq:associative}
\end{equation}
which follow from associativity of the product $L_{g_i}L_{g_j}L_{g_k}$.
Now, two states are in the same phase if their factor sets are equivalent.
That is, they are in the same phase whenever their factor sets can
be transformed
into one another by making
a ``gauge transformation" on the effective symmetry operators of
one of the states, $L_{g_i}\rightarrow e^{i\alpha(g_i)}L_{g_i}$.
Thus the two sets of phases
\begin{equation}
\rho'(g_i,g_j)=\rho(g_i,g_j)+\alpha(g_i)+\alpha(g_j)-\alpha(g_ig_j)
\label{eq:gauge}
\end{equation}
 describe the same phase. When Eq. (\ref{eq:associative}) is solved and the solutions
are broken into distinct classes according to Eq. (\ref{eq:gauge}),
then there are a finite number of distinct classes.
The group structure of the phases is easy to see in this result; a pair
of side-by-side chains whose factor sets are $\rho_1$ and $\rho_2$ has
a combined factor set of $\rho_1+\rho_2$.
With this operation, the set of classes for a given symmetry group $G$ also becomes a group, which is called the
second cohomology group $H^2(G)$, or the
``Schur multiplier of $G$"\cite{Schur}.

Note that \emph{any} state with a nontrivial $\rho$ that cannot be reduced
to zero must have edge states.  Indeed, if the ground state is unique,
it is an eigenfunction of all the $L_g$'s, with eigenvalue $e^{-i\alpha(g)}$ say.
Eq. (\ref{eq:factorset}) then says that $\rho(g_i,g_j)$ can be gauged to zero.

The minimal symmetry that can be used to define topological phases is
time reversal, where we assume that $T^2$=1 (i.e., the particles are spinless).
$T$ is antiunitary, so $L_T$ is as well.  It squares to a pure phase;
however the phase cannot be absorbed for an antiunitary operator, and has to be $+1$
or $-1$, giving two phases. In the latter case, the edge state forms a doublet by Kramers's theorem.
Thus even particles which individually have no ``spin degrees of freedom," i.e.
ones whose energy levels are all singlets, can form a chain with edge states!

Topological phases might occur in insulating systems
of hopping bosons as well as in spin chains; a state like this would
be a topological variant on a Mott insulator. An example of this type
was proposed by \cite{Berg}, but this state requires inversion symmetry, so
there are no edge states.
Further work is needed to find realistic topological phases besides the antiferromagnetic spin  1 chain.

\begin{table*}
\centering
\begin{ruledtabular}
\begin{tabular}{|l|c|l|}
  {\em Symmetry} & \em Topological Classification & {\em Comments} \\
  \hline
   No symmetry & 0  & Topological states require symmetry in 1D.\\
  $SO_3$ & $\mbz_2$ & Spin chain with \emph{integer} spins.\\
  $Z_2^T$ such that $T^2=1$. & $\mbz_2$ & Time reversal symmetry \\
  $U(1)$ & 0 & Charge conserved. \\
  $\mathbb{Z}_n$ &0& \\
  $\mathbb{Z}_n\times\mathbb{Z}_m$ & $\mathbb{Z}_{(n,m)}$ & $(a,b)\equiv$\ greatest
common divisor of $a$ and $b$.\\
  $U(1)\rtimes \mathbb{Z}^{particle-hole}_2$ & $\mathbb{Z}_2$ & Particle-hole symmetry\\
Fermions; $\mbz_2^{charge}$ & $\mathbb{Z}_2$ & Superconductors\\
Fermions; $\mbz_2^{charge}\times \mbz_2^T$; $T^2=1$ & $\mathbb{Z}_8$ & Superconductor
with time reversal of spinless fermions.\\
  \hline
\end{tabular}
\caption{Classification of some gapped D=1+1 dimensional phases of bosons and
fermions.}
\label{table2}
\end{ruledtabular}
\end{table*}

\emph{Topological Phases of Fermions: Interacting Version of Topological Insulators}--The table just found for bosons is the analogue of
the standard tables of topological insulators when interactions
are taken into account.
Each of the 10 symmetry classes of topological insulators corresponds
to a particular symmetry group.  For example, the class $A$ (the quantum
Hall case, in two-dimensions) is usually said not to have any symmetry,
because the matrices describing the one-body equation of motion
 do not have any symmetry.
But this system does have particle number conservation which is an important
symmetry for classifying this state (the Hall conductivity is the amount
of charge transported).  Thus the symmetry
group is $U(1)$.
Any superconducting phase, on the other hand, does not have a full $U(1)$
symmetry, but only $\mathbb{Z}^{charge}_2$ symmetry: the term $\Delta \psi_\uparrow\psi_\downarrow$ that describes pairing (at least in mean-field theory) breaks $U(1)$ symmetry.
Thus, the class DIII is described\footnote{In fact there are really at least two ways to interpret each of the D and C
symmetry classes when interactions start to play a role. In the non-interacting
case, this can describe either a system with
an actual particle-hole symmetry or a superconductor, where the particle-hole
symmetry is a mathematical artifact.  But in the interacting case, these
two interpretations correspond to different symmetry groups.}
 by $\mathbb{Z}_2^T\times\mathbb{Z}_2^{charge}$, where $\mathbb{Z}_2^T$ refers to
time reversal. Note that we model
a superconductor by explicitly inserting a pairing term into the Hamiltonian
to ensure that the chain is gapped. We do not consider
the effects of phase fluctuations.

Besides these 10 Altland-Zirnbauer symmetry groups, the system can
have internal symmetries. Unlike in the interacting case,
each possible symmetry group has to be treated separately.
For the non-interacting case, only the symmetries that are non-unitary
(or flip the sign of the energy) need to be classified, explaining
why the Altland-Zirnbauer classes are so important.  Any ordinary internal symmetry
commutes with the momentum, so it can be used to divide the band
structure up into blocks according to the states' quantum numbers.
Then each block
may be classified according to the Altland-Zirnbauer classification\footnote{A block may be classified by a different Altland-Zirnbauer than the system as a whole, if, e.g. the action of the internal symmetry on the block is described by a complex representation. Ref. \cite{Dyson} explains how this works.}.

On the other hand, in an interacting system, interactions exist between the different blocks,
and hence
all symmetries (not just the antiunitary ones) can become fractionalized by interactions; there are more (infinitely many more) than ten classes in the presence of interactions.

The methods of the previous section can be applied to fermions as well with a small modification;
this gives a methodical explanation of the $\mathbb{Z}_8$ classification of
superconductors of the type $BDI$\cite{Fermions}.
Two special features of fermions have to be taken into account.
These features already
protect two phases even
when no symmetry is imposed.
The first feature is that there is no way to break the conservation of
fermion parity $Q$ (defined as $\pm 1$ depending on whether
there is an odd or an even number of fermions)
because unpaired fermions cannot form a condensate.  This
is just a single $\mathbb{Z}_2$ symmetry, and for bosons that
would not be enough to protect distinct phases.
However, the second key feature is that the factors $L_Q$ and $R_Q$ of $Q$
may either \emph{commute} or \emph{anticommute} because they can contain fermionic
creation and annihilation operators.  These correspond to distinct phases.

When $L_Q$ and $R_Q$ anticommute,
the consequence is that there is a fermion state that is split
between the two ends of the chain. The commutation
properties of $L_Q$ and $R_Q$ show that there is a $2$-fold degenerate ground state
associated with the system as a whole, not with the separate ends. This represents
the two states (occupied and not occupied) associated with a fermion. If there is an
observer at each end, then either one of them can switch the system between
these two states, by applying $L_Q$ or $R_Q$. On the
other hand \emph{neither} observer acting alone can measure which of the two states the system is in (as
$L_Q,R_Q$ are fermionic, they are not observable).  This two-level
system can therefore act as a robust q-bit because it cannot decohere
through ``observation" by the environment\cite{kitaevMajorana}.

This classification into two phases agrees with the appropriate
non-interacting classification, i.e., the
type $D$ topological insulators.  In fact, these are classified into two phases,
depending on whether there is an even or an odd number of Majorana operators $a_{R/Li}$
(satisfying $a_{R/Li}^2=1$) at an end.
The  total parity of a state below the gap in such a non-interacting
insulator
is given by the product over all Majorana and ordinary
fermion edge modes: $\prod_i a_{Li}\prod_i a_{Ri} (-1)^{\sum_i f_{Li}^\dagger f_{Li} +\sum f_{Ri}^\dagger f_{Ri}}$ (up to a phase). The $L_Q$ and $R_Q$ operators are
the parts corresponding to the two ends.  Thus $L_Q$ is fermionic in a phase
with an odd number of Majorana modes at each end. The arguments above show that
a two-fold degeneracy survives in this case even when the Majorana modes interact.

When an additional symmetry is added, the interacting and non-interacting
cases behave differently. If we consider ``spinless time reversal symmetry", i.e.,
$T^2=\mathds{1}$, then the classification table for class $BDI$ says
that the phases are classified by $\mathbb{Z}$.  However, it seems unlikely
to be possible to get a classification into an infinite number of states
by studying the symmetries of the edge states,
since each symmetry or set of symmetries
gives rise to only a few discrete phases like $\phi$ in Eq. (\ref{eq:PHI}).
In fact, analyzing the symmetries shows that there are just 8 interacting phases.
They are classified by the binary choice
just described (whether $L_Q$ and $R_Q$ are bosonic
or fermionic) together with one that describes whether
$L_{T}^2=\pm 1$ and another that describes whether $L_T$ and $Q_T$
commute or anticommute, giving $8$ phases in all. One can also see this starting from the noninteracting limit where any integer number of Majorana zero modes at the edge may be realized.  Suppose $k$ identical chains each with a Majorana mode at the end
are placed side by side. Say the modes are all even under time-reversal.
Then these cannot couple to one another in quadratic order because
of time reversal symmetry (a coupling such as $a_1a_2$ isn't Hermitian, while
$ia_1a_2$ violates time-reversal symmetry) so a degeneracy of $2^k$ for the
entire segment is protected.  However, if
there are interactions, there can be a coupling of $4$ Majoranas ($a_1a_2a_3a_4$).
This
reduces the degeneracy \emph{per end} from $4$-fold to $2$-fold. That suggests that
when there are enough Majoranas at the ends of the chain, it should
be possible to couple them in a way that removes all the degeneracy.  And in fact
that is possible once there are $8$ Majoranas.  The 2-fold degeneracy of an end of
the $k=4$
state
is protected by time reversal.
The doublet is a Kramers doublet (although $T^2=1$ in
the microscopic degrees of freedom, $T^2=-1$ in the fractionalized edge states).
It can be described as a spin-$\frac{1}{2}$ (though not with rotational
symmetry). When two sets of four chains are coupled together, the spin-$\frac{1}{2}$
in each can form a singlet,
and the edge degeneracy is removed.

\emph{Summary} -- In one dimension, distinct phases exist only when the Hamiltonian is symmetric, with the exception of Majorana fermions (these are also related to a
symmetry, $Q$, but \emph{this} symmetry is automatic). These phases have
fractionalized edge states, and the states can be
characterized based on the factor set for the projective transformations
of the edge states. These phases form a group when addition is defined as placing two states side-by-side.

\subsection{Topological SRE Phases in Higher Dimensions}
 Now we turn to two dimensions. In 2D topological order, as exemplified
by the fractional quantum Hall state,
becomes possible for the first time. In contrast with topological insulators,
as exemplified by the integer quantum Hall and quantum spin Hall states, these systems
have non-trivial bulk excitations with anyon statistics.
Many instances of topologically ordered states have been studied, in connection with the fractional quantum Hall effect and quantum spin liquids. The surprise in recent times has been the discovery of interacting topological phases with SRE \cite{KitaevUnpublished,Chencoho2011} in $d>1$, the analogue of the \emph{integer}
quantum Hall effect for interacting systems.  Hence we will focus on these phases.
The notable
feature of topological insulators is their surface states, and the SRE
phases share this property with them: there are no non-trivial excitations in their
bulk. In contrast, topologically ordered states do not necessarily have edge
states (although some do).  We will specialize to gapped phases built out of {\em bosons}. This ensures that interactions are essential to the description (since non-interacting bosons can only form superfluids), so that we will not mistakenly reproduce a documented free fermion topological insulator. 

The most basic distinction between phases arises in the absence of any symmetry. For one dimensional gapped phases, there are no nontrivial distinctions in the absence of symmetry. However, in two dimensions, this is no longer true, as pointed out by Kitaev.  A convenient way to distinguish topological phases is via quantized response. While the best known quantized response is  Hall conductance, this implicitly assumes the presence of a conserved electric charge that couples to the applied electric field. In the absence of symmetry, there is no conserved charge. However, one can still define a quantized thermal Hall conductance $\kappa_{xy}$. Its ratio to
the temperature $\frac{\kappa_{xy}}{T}$ is quantized in units of $L_0=\frac{\pi^2k_B^2}{3h}$.  Since energy is always conserved, this phase remains distinct even when charge conservation is violated.
It is well known that in the context of free fermion topological phases, the chiral superconductors that break time reversal symmetry (class D) may be distinguished by the number of Majorana modes at the edge, with thermal Hall conductance quantized, in units of $\kappa_{xy}=\frac12L_0T$ while the integer quantum Hall phases have Hall conductances quantized in units of $L_0$. Note, as usual this is sharply defined only in the limit of zero temperature ($T\rightarrow 0$), and the quantized response is actually the ratio of thermal Hall conductance to temperature.  This quantized thermal conductance can also be defined for interacting phases. Kitaev showed that it is possible to construct a bosonic insulator (which is necessarily interacting) with chiral edge modes that has a quantized thermal Hall conductance.  Moreover, assuming there is no bulk topological order, the thermal Hall conductance $\kappa_{xy}/T$ of such an insulator must be quantized to multiples of $8L_0$ (16 times the minimum conductance of an SRE system of fermions). When the bulk has topological order, the thermal Hall conductance can be a fraction of this, similar to the ordinary Hall conductance in a fractional quantum Hall state.

 One can also discuss topological phases protected by a symmetry - for example U(1) charge conservation. This leads to an integer classification of SRE phases \cite{Chencoho2011,LuAv,SenthilLevin} of gapped bosons in 2D. Just as in the previous case, the topological phases are distinguished by a quantized response, which in this case is quantized Hall conductance. Interestingly, for bosons, the Hall conductances are quantized to {\em even} integers, i.e. $\sigma_{xy}=2n\frac{q^2}{h}$\cite{LuAv}, assuming no topological order.
While there are different ways to see these results, the easiest is via the Chern Simons K-Matrix formulation which we briefly describe.

{\bf $K$-matrix formulation:} In the $K$ matrix formulation of quantum Hall states, a symmetric integer matrix appears in the Chern-Simons action:
($\hbar=1$, and summation is implied over repeated indices $\mu,\nu,\lambda=0,1,2$):
\begin{eqnarray}\label{cs action}
&4\pi\mathcal{S}_{CS}=\int d^2xdt~\sum_{I,J}{\epsilon_{\mu\nu\lambda}}a^I_\mu [{\bf K}]_{I,J}\partial_\nu a_\lambda^J
\end{eqnarray}
While this has been utilized to discuss quantum Hall states with Abelian topological order, it is also a powerful tool to discuss topological phases in the {\em absence} of topological order. The latter requires $|det{\bf K}|=1$ (i.e. $\bf K$ is a {\em unimodular} matrix). Furthermore, in order that all excitations are bosonic, we demand that the diagonal entries of the matrix are {\em even} integers. The bulk action also determines topological properties of the edge states. For example, the signature of the $K$ matrix (number of positive minus negative eigenvalues) is the chirality of edge states - the imbalance between the number of right and left moving edge modes. Maximally chiral states have all edge modes moving in the same direction. Physically, the fluxes $j^{\lambda}_I=\epsilon^{\mu\nu\lambda} \partial_\mu a^I_\nu$ are often related to densities and currents of bosons of different flavors.
\begin{figure}
 \includegraphics[width=0.45\textwidth]{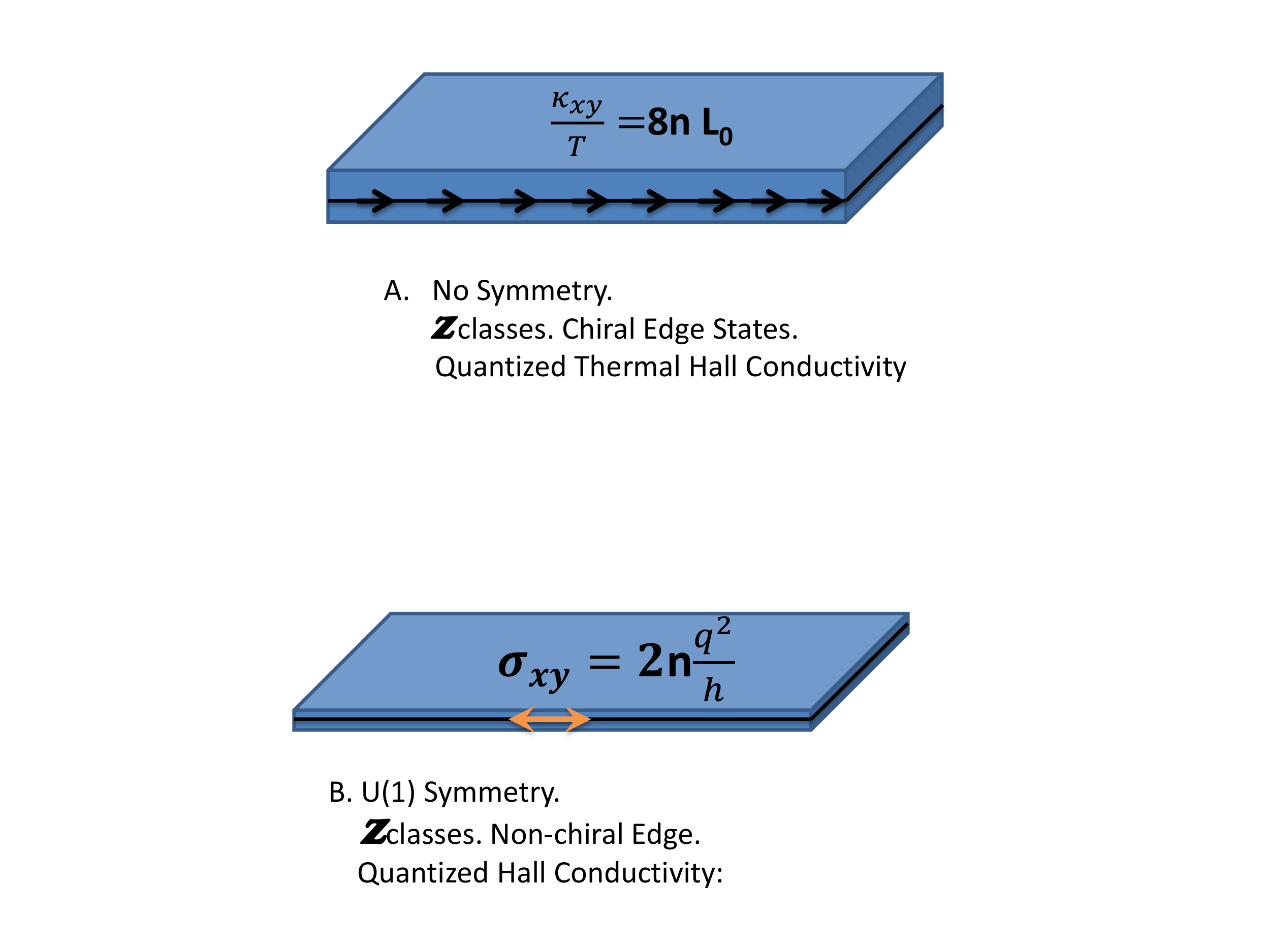}
\caption{Summary of some simple `integer' bosonic topological phases. 1) A chiral phase of bosons (no symmetry required). An integer multiple of eight chiral bosons at the edge is needed to evade topological order, leading to a quantized thermal Hall conductance $\kappa_{xy}/T=8nL_0$ in units of  the universal thermal conductance $L_0=\frac{\pi^2k_B^2}{3h}$. These are bosonic analogs of chiral superconductors. (2) A non-chiral phase of bosons protected by $U(1)$ symmetry (eg. charge conservation). Distinct phases can be labeled by the quantized Hall conductance $\sigma_{xy}=2n\sigma_0$, which are even integer multiples of the universal conductance $\sigma_0=q^2/h$ for particles with charge $q$. These are bosonic analogs of the integer quantum Hall phases.}\label{fig:summary}
\end{figure}

We begin by looking for maximally chiral states of bosons without topological order. These are bosonic analogs of the integer quantum Hall effect or chiral superconductors of fermions; they have a \emph{thermal} Hall coefficient of $nL_0T$ where $n$
is the dimension of the $K$ matrix. The smallest dimension of bosonic unimodular $K$ matrix that yields a maximally chiral state is 8\cite{LuAv}. This is consistent with the prediction of Kitaev, derived from topological field theory.
An example of such a $K$ matrix is given by:

\begin{equation}
K^{E_8}=\left(
    \begin{array}{cccccccc}
      2 & -1 & 0 & 0 & 0 & 0 & 0 & 0\\
      -1 & 2 & -1 & 0 & 0 & 0 & 0 & 0 \\
      0 & -1 & 2 & -1 & 0 & 0 & 0 & -1 \\
      0 & 0 & -1 & 2 & -1 & 0 & 0 & 0 \\
      0 & 0 & 0 & -1 & 2 & -1 & 0 & 0 \\
      0 & 0 & 0 & 0 & -1 & 2 & -1 & 0 \\
      0 & 0 & 0 & 0 & 0 & -1 & 2 & 0 \\
      0 & 0 & -1 & 0 & 0 & 0 & 0 & 2 \\
    \end{array}
  \right)
  \label{E8}
\end{equation}
This $K$ matrix is the Cartan matrix of the group $E_8$, and hence we call this phase the Kitaev E$_8$ phase; see Fig. \ref{fig:summary}a.  It has edge modes protected by the thermal Hall effect, which does not require any symmetry.

We now consider non-chiral states of bosons, with equal number of left and right moving edge modes, in order to investigate other routes to protected edge modes. In the absence of symmetry we believe there are no nonchiral topological phases with $|\det{\bf K}|=1$. However, the presence of a symmetry can lead to new topological phases. The integer quantum Hall states of bosons are in fact modeled by the very simple $K$ matrix:
$
K=
\left(
\begin{array}{cc}
 0   & 1   \\
  1 & 0
\end{array}
\right).
$
Due to the presence of a conserved charge, the electromagnetic response of the system can be computed by coupling an external gauge field $A_\mu$.  The field theory then is:
\begin{equation}
{\mathcal L}= \frac1{2\pi} \epsilon^{\mu\nu\lambda}\left [ a_{1\mu}\partial_\nu a_{2\lambda} -  (a_{1\mu}+a_{2\mu})\partial_\nu A_\lambda\right ]
\label{U(1)}
\end{equation}
where we have taken the coupling to have the simplest interesting form\footnote{The coupling between $A_\mu$ and $a_{j\mu}$ must have an integer coefficient; $\mathrm{curl}\ a_j$ represents a density of bosons of some type, which must have integer charges. We set both coefficients equal to one since this gives the smallest non-zero Hall conductivity.}.
Integrating out  the internal fields gives the effective action for the electromagnetic response:
\begin{equation}
{\mathcal L}_{em}= -\frac1{2\pi} \epsilon^{\mu\nu\lambda}A_\mu \partial_\nu A_\lambda ,
\end{equation}
which corresponds to a $\sigma_{xy}=2$. 
A pictorial representation of this phase is given in Figure \ref{fig:summary}a. Intuitively, the non-chiral modes  at the edge implied by  Eq.\ (\ref{U(1)}) can be viewed as oppositely propagating modes, one of which is charged and the other neutral\cite{LuAv,WenSU(2),SenthilLevin}. This prohibits backscattering between the modes. Another viewpoint is that it is a regular Luttinger liquid except in that it transforms under the internal $U$(1) symmetry in a way that is impossible to realize in a 1D Luttinger liquid\cite{LuAv,WenSU(2)}. For example, in this particular case, both the conjugate fields $\theta,\phi$ of the Luttinger liquid transform like charge phases under the $U(1)$ symmetry. Recent work has argued that the Integer Quantum Hall state of bosons could appear in interacting systems of bosons in the lowest Landau level\cite{SenthilLevin}, that could be realized in cold atom systems exposed to an artificial gauge field.

In the presence of more complicated symmetries, there are many more topological phases.  These phases are at least partly classified by the third cohomology
group of $G$, as shown by Ref. \cite{Chencoho2011, ChenEdge}. The classification is still not entirely complete, especially when Majorana edge modes are present.
Besides producing new phases in bosonic systems, interactions
can also reduce the number of topological insulating phases in fermionic
systems in higher dimensions (similar to the one-dimensional case)\cite{Gu,2dFermions,LuAv}.

{\bf SRE Phases and Topological Order:}
There are interesting and deep relations between SRE topological phases and states with topological order. Levin and Gu\cite{LevinGu} pointed out that the relation can be viewed as a duality. It is well known that the 2+1D quantum Ising model with Z$_2$ global symmetry is dual to a Z$_2$ gauge theory, with the disordered phase mapping to the deconfined phase of the gauge theory. However, there exists a SRE topological phase with Z$_2$ symmetry, distinct from the disordered phase of the Ising model. Under the same duality it maps to a twisted version of the Z$_2$ gauge theory, the doubled semion model. This approach can be used to give an alternative interpretation of the classification
obtained by Refs. \onlinecite{Chencoho2011,ChenEdge}.
The idea is to study the properties of the gauge excitations\cite{unpublished}, point-like excitations similar to vortices in a superconductor that are created by passing a flux line through the system for an element $g_i$ of the symmetry group.
If we take three such gauge-fluxes $g_1$, $g_2$, $g_3$ arrayed along a line\footnote{Arranging them on a line allows us to label fluxes by group elements rather than congugacy classes.} then they  can be combined
to make an excitation with flux $g_1g_2g_3$.  This can be done in two orders, fusing $1,\,2$ first and then $3$ or $2,\,3$ first and then $1$.
The difference between the two orders gives a phase factor $e^{i\kappa(g_1,g_2,g_3)}$.  This phase factor is ambiguous because a phase can
accumulate as two of the fluxes are brought together, depending on arbitrary choices one makes about how they are moved. Nevertheless there are still inequivalent
choices of $\kappa$'s, similar to the inequivalent classes of $\rho$'s in
the one-dimensional case.
This gives
a way to distinguish phases.  The group of $\kappa$'s is called the third
cohomology group, which therefore classifies the bosonic states (at least partly)
as Ref. \cite{Chencoho2011,ChenEdge} found.
A similar result about discrete gauge theories has been known\cite{DWitten}, that there can be extra phase factors in the mutual statistics and fusion
of the anyons for exactly this reason.

\subsection{Edge States and Topological Order}
We have been characterizing SRE phases by their edge states.   However edge states can exist in fractional quantum Hall states as well.  Is there a connection between the fractional edge states and the existence of anyons?  After all,
anyons and edge states both have a handedness. In fact, the thermal Hall
conductance of the edge of a system of bosons is \emph{partly} determined by the collection of anyons
through the following strange relation\cite{ThermalAnyons}:
\begin{equation}
e^{\frac{\pi i}{4}\left[\frac{\kappa}{TL_0}\right]}=\frac{\sum_{i=1}^{N_A} e^{2\pi i s_i}}{\sqrt{N_A}}
\end{equation}
where the sum is over the $N_A$ species of anyons and $2\pi s_i$ are their
statistical angles (and we have also assumed that the phase is abelian).
Thus systems with fractional values of $\frac{\kappa}{8 TL_0}$ must have
anyons (explaining why the Kitaev $E_8$ state has $\kappa=8TL_0$). Conversely, the
set of anyons determines $\frac{\kappa}{TL_0}$ partly; it must have the form $\frac{p}{q}+8n$ where $\frac{p}{q}$ is a certain fraction and $n$ is an indeterminate integer. If this
is not a multiple of $8$, the edge states are protected without assuming
any symmetry.

A related subject that has received much attention is the interplay between topological order and surface states in the presence of global symmetries. Perhaps most relevant to this article  is the question of stability of edge states in fractional topological insulators, in the presence of time reversal symmetry. A general condition for abelian states was worked out in \cite{LevinStern}. Similar considerations in the context of fractionalized three dimensional phases with time reversal symmetry \cite{PesinBalents,FractionalTopological} were discussed. In insulators that are adiabatically connected to free fermions, the magneto electric polarizability $\theta$, is fixed to be multiples of $\pi$ to preserve time reversal symmetry. In Refs. \onlinecite{FractionalTopological} it was argued that fractional values of $\theta/\pi$ are compatible with time reversal symmetry only if the bulk possesses topological order.

\subsection{Outlook}
There are several open questions in this rapidly evolving field. Perhaps most urgently, the physical nature of SRE topological phases in $d>1$ need to be more deeply understood. Some recent work has begun to discuss properties of 3D SRE topological phases\cite{SenthilAV,Swingle} predicted in Ref. [\onlinecite{Chencoho2011}]. Another important open question concerns the physical realization of the bosonic topological phases in quantum magnets and lattice cold atom systems, and the types of interactions required to realize them in various dimensions. It also remains to be seen if interacting fermions can lead to new topological phases.
So far, all the cases of fermionic topological states that have been examined\cite{Gu,LuAv} either reduce to bosonic phases formed from bound states of the fermions or to non-interacting topological insulators. The deep connections between topological order and SRE topological phases, as well as the interplay of symmetry and topological order remain to be more deeply explored.

{\bf Acknowledgements:} We would like to acknowledge insightful discussions with Alexei Kitaev, Lukasz Fidkowski and Leon Balents and acknowledge our collaborators on related projects Sergey Savrasov, Xiangang Wan, Yuan-Ming Lu, T. Senthil, Tarun Grover, Frank Pollman, Erez Berg and Masaki Oshikawa. This work was upported in part by Grant Number NSF DMR-1206728.

\end{document}